\documentclass[conference]{IEEEtran}
\IEEEoverridecommandlockouts
\usepackage{cite}
\usepackage{amsmath,amssymb,amsfonts}
\usepackage{algorithmic}
\usepackage{graphicx}
\usepackage{textcomp}
\usepackage{xcolor}
\usepackage{multirow}
\usepackage{caption}
\usepackage{enumerate}
\usepackage{booktabs}

\def\BibTeX{{\rm B\kern-.05em{\sc i\kern-.025em b}\kern-.08em
    T\kern-.1667em\lower.7ex\hbox{E}\kern-.125emX}}
\begin{document}

\title{Channel Adaptation for Speaker Verification Using Optimal Transport with Pseudo Label \\
\thanks{This work was supported by National Key R\&D Program of China (No.2023YFB2603902) and Tianjin Science and Technology Program (No. 21JCZXJC00190).}
}

\author{
    \IEEEauthorblockN{
        Wenhao Yang\IEEEauthorrefmark{1}, 
        Jianguo Wei\IEEEauthorrefmark{1}, 
        Wenhuan Lu\IEEEauthorrefmark{1}, 
        Lei Li\IEEEauthorrefmark{2}\IEEEauthorrefmark{3} and
        Xugang Lu\IEEEauthorrefmark{4}
    }
    \IEEEauthorblockA{
        \IEEEauthorrefmark{1}College of Intelligence and Computing, Tianjin University, Tianjin, China\\
        \IEEEauthorrefmark{2}University of Washington 
        \IEEEauthorrefmark{3}University of Copenhagen\\
        \IEEEauthorrefmark{4}National Institute of Information and Communications Technology, Japan\\
    }
}

\maketitle

\begin{abstract}

Domain gap often degrades the performance of speaker verification (SV) systems when the statistical distributions of training data and real-world test speech are mismatched. Channel variation, a primary factor causing this gap, is less addressed than other issues (e.g., noise).  Although various domain adaptation algorithms could be applied to handle this domain gap problem, most algorithms could not take the complex distribution structure in domain alignment with discriminative learning. In this paper, we propose a novel unsupervised domain adaptation method, i.e., Joint Partial Optimal Transport with Pseudo Label (JPOT-PL), to alleviate the channel mismatch problem. Leveraging the geometric-aware distance metric of optimal transport in distribution alignment, we further design a pseudo label-based discriminative learning where the pseudo label can be regarded as a new type of soft speaker label derived from the optimal coupling. With the JPOT-PL, we carry out experiments on the SV channel adaptation task with VoxCeleb as the basis corpus. Experiments show our method reduces EER by over 10\% compared with several state-of-the-art channel adaptation algorithms.

\end{abstract}


\section{Introduction}
Speaker verification (SV) is the task of comparing the speaker's identities in two utterances. Current state-of-the-art speaker verification techniques adopt the x-vectors \cite{snyder2018x} framework. First, the speaker embedding extractor is trained for speaker classification. Then, test embeddings extracted from the testing speech are compared using the backend score module, like cosine similarity and PLDA \cite{ioffe2006probabilistic}.

In real-world scenarios, a persistent domain gap exists between training and testing data that degrades the SV performance \cite{garcia2014supervised,chen2021self,9794604}. A key source of this gap is channel variation, which arises from recording devices and transmission methods \cite{sadjadi20202019,sadjadi20222021}. The variation results in changes in speech encoding \cite{6926067,qin2022robust,yang2024robust}, and leads to discrepancies in data distribution. Assuming training data as the source domain and test data as the target domain, adapting models trained on the source domain to the target domain can be accomplished using domain adaptation (DA). 

Unsupervised Domain Adaptation (UDA), which assumes no speaker labels are available in the target domain, is more relevant to real-world testing scenarios than supervised domain adaptation. UDA methods, which aim to minimize the distribution gap between domains, predefined statistical distances between two distributions, including Correlation Alignment (CORAL) \cite{sun2016return} and its variants CORAl++ \cite{li2022coral++} and DeepCORAL \cite{sun2016deep}, Wasserstein Distance \cite{courty2016optimal} and Maximum Mean Discrepancy (MMD) \cite{borgwardt2006integrating,lin2020multi}. Another way of domain alignment is by using domain adversarial training, including Domain-Adversarial Neural Networks (DANN) \cite{ganin2016domain,li2020contrastive,zhang2022learning}, to learn representations that are invariant to domain discriminators. It is known that target domain discriminative learning is also important besides domain alignment in domain adaptation. Fewer algorithms focus on target domain discriminative learning, including contrastive learning \cite{li2020contrastive,chen2021self} and clustering \cite{mao2023cluster}, as a complement to alignment. These methods do not account for the complex distribution structures in adaptation and thus do not perform well in SV tasks with complex scenarios due to channel variations.

UDA methods in speaker verification primarily focused on noise conditions, overlooking other intrinsic inter-speaker and intra-speaker differences between source and target data \cite{sun2016deep,lin2020multi,li2022coral++,zhang2022learning}. These methods often neglect the overlap of speaker classes and do not consider partial domain adaptation. Aligning the distribution of a subset of speaker classes with the overall speaker distribution without maintaining discriminative properties can lead to over-alignment. Furthermore, apart from background noise (additive or convolutional), speech signals may vary due to bandwidth differences in transmission channels. Consequently, speaker verification models may make wrong decisions in determining whether two speech signals are from the same speaker \cite{yang2024robust}.

\begin{figure*}[tb]
\centering
\includegraphics[width=0.75\linewidth]{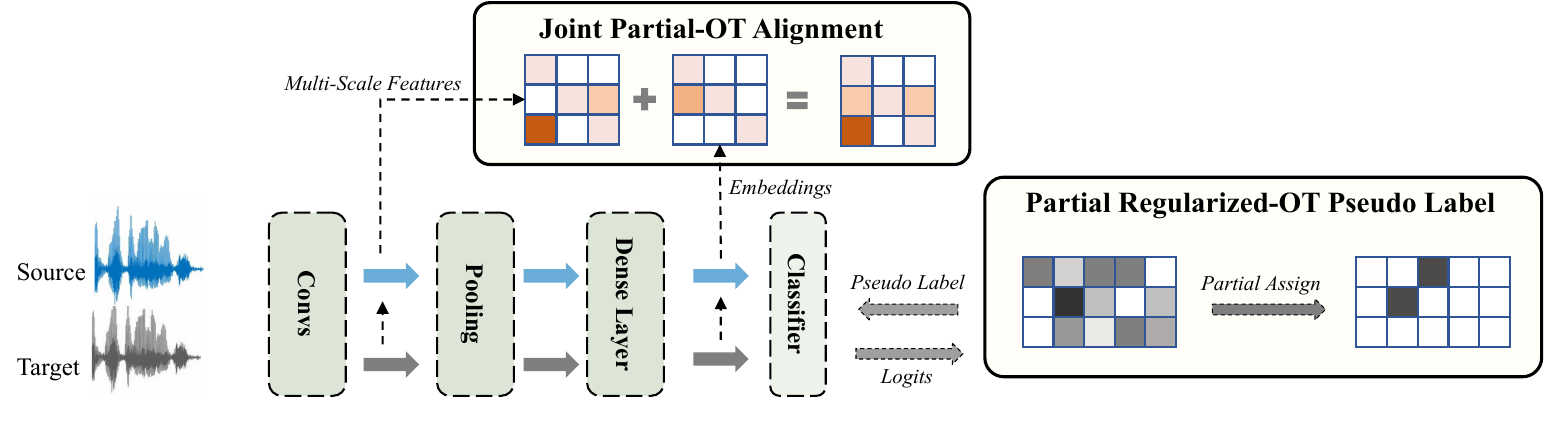}
\caption{Channel Domain Adaptation for Speaker Verification with Joint Partial Optimal Transport Alignment and Pseudo Lable Discriminative Learning}
\label{fig:jpot}
\end{figure*}

Optimal transport (OT), a mathematical method for finding an efficient way to transport one probability distribution to another, can serve as a natural measure of discrepancy between two probability distributions \cite{fournier2015rate,flamary2021pot}. As a geometric-aware distance metric for data distributions, OT has demonstrated its superiority and efficiency in various tasks, including image processing, classification, and segmentation \cite{courty2016optimal,courty2017joint,xiao2021dynamic}. Additionally, advances in OT include utilizing DeepJDOT for joint distribution OT in neural networks \cite{damodaran2018deepjdot}, partially aligning distribution for language identification \cite{lu2021unsupervised}, and efficiently pseudo-labeling or clustering for classification \cite{taherkhani2021self,liu2023cot}. 

Building on these insights, we propose a unified OT-based domain adaptation method that performs partial domain alignment and pseudo-labeling for discriminative learning for channel adaptation in speaker verification. We term this method Joint Partial Optimal Transport with Pseudo Label (JPOT-PL). The contributions of this paper are as follows:

\begin{enumerate}
\itemsep=0pt
\item We explored Optimal Transport with joint distribution for channel adaptation in speaker verification.
\item We propose combining pseudo-label learning to enhance the discriminative ability within unsupervised domain adaptation for channel variability in speech signals.
\item We present a comparative study on channel adaptation for the speaker verification task.
\end{enumerate}

\section{Methodology}
\label{sec:format}

In this section, we introduce two modules in JPOT-PL: Joint Partial-OT Alignment and Partial Regularized-OT Pseudo Label for speaker channel domain adaptation as in Figure~\ref{fig:jpot}. 

\subsection{Preliminary}

In the deep speaker verification, the whole framework could be defined as:
\begin{equation}\label{eq2}
y = g \circ h(x); \textbf{e} = h(x) 
\end{equation}
where $x$ is the input speech signals, $h(x) $ is the speaker embedding extractor that extracts speaker embeddings $\textbf{e}$ from speech $x$ and $g$ is the classifier. In domain adaptation, there is a labeled source domain $D^s$ and an unlabeled target domain $D^t$. The adaptation method tries to adapt model $g \circ h(x)$ from $D^s$ to $D^t$. So the embeddings $e^t$ of $x_1$ in $D^t$ are discriminative as the embeddings $e^s$ of $x_2$ in $D^t$. 

Optimal Transport \cite{courty2016optimal} transports one probability distribution to another with minimum cost. Formally, OT searches for a probabilistic coupling $\gamma \in (u_1, u_2)$ (transport plan) between two distributions $u_1$ for $D^s$ and $u_2$ for $D^t$ that minimizes the cost:
\begin{equation}\label{eq2}
\begin{split}
OT(u_1, u_2) \triangleq \min_{(x_1, x_2) \in(u_1, u_2)} \sum {C(x_1, x_2) \gamma(x_1, x_2)}
\end{split}
\end{equation}
where $C(x_1, x_2)$ is the cost function measuring the distance between samples $x_1$ and $x_2$. The $OT$ measures the discrepancy between source and target distributions. To transfer the model into the target domain, minimizing the OT distance also known as Wasserstein Distance, is equal to the alignment of these distributions.

\subsection{Joint Partial-OT Alignment}

\begin{table*}[h]
\centering
\caption{Performance ( EER($\%$) ) of degraded channel speech for speaker verification. [0,1,2,5]: noise volts in radio channels.}
\begin{tabular}{lcrrrrrrrrrrrr}
\toprule
\multicolumn{1}{c|}{\multirow{2}{*}{\textbf{Adaptation}}} & \multicolumn{1}{c|}{\multirow{2}{*}{\textbf{Source}}} & \multicolumn{4}{c|}{\textbf{64 Spks}}                                          & \multicolumn{4}{c|}{\textbf{256 Spks}}                                         & \multicolumn{4}{c}{\textbf{All Spks}}                                         \\
\multicolumn{1}{c|}{}            &     \multicolumn{1}{c|}{}                     & \multicolumn{1}{c}{0} & \multicolumn{1}{c}{1} & \multicolumn{1}{c}{2}     & \multicolumn{1}{c|}{5} & \multicolumn{1}{c}{0} & \multicolumn{1}{c}{1} & \multicolumn{1}{c}{2}     & \multicolumn{1}{c|}{5} & \multicolumn{1}{c}{0} & \multicolumn{1}{c}{1} & \multicolumn{1}{c}{2}     & \multicolumn{1}{c}{5} \\
\midrule
No Adaptation & 2.32 &  8.27 & 12.65 & 16.89 & 38.68  &  8.27 & 12.65 & 16.89 & 38.68 &  8.27 & 12.65 & 16.89 & 38.68 \\
\midrule
DANN \cite{ganin2016domain} &  2.29 &  7.10 & 10.53 & 14.47 & 37.45 & 7.18 & 10.48 & 14.44 & 37.42 & 7.13 & 10.42 & 14.36 & 37.41  \\
DeepCORAL \cite{sun2016deep} &  2.34 &  8.41 & 12.76 & 16.80 & 37.54 & 8.44 & 12.53 & 16.55 & 37.74 & 8.52 & 12.57 & 16.59 & 37.86  \\
MMD \cite{borgwardt2006integrating} &  2.31 &  7.03 & 11.23 & 15.42 & 39.12 & 7.05 & 11.18 & 15.28 & 38.98 & 7.03 & 11.06 & 15.22 & 38.99  \\
OT \cite{courty2016optimal} &  2.30 &  7.76 & 11.44 & 15.32 & 37.66 & 7.51 & 10.91 & 14.78 & 37.57 & 7.41 & 10.78 & 14.69 & 37.59  \\
DeepJDOT \cite{damodaran2018deepjdot} &  \textbf{2.28} &  6.92 & 10.49 & 14.47 & 37.86 & 7.16 & 10.54 & 14.45 & 37.80 & 7.10 & 10.45 & 14.34 & 37.80  \\
\textbf{JPOT-PL (Ours)}  &  \textbf{2.28} &  \textbf{6.73} & \textbf{9.84} & \textbf{13.67} & \textbf{36.21} & \textbf{6.37} & \textbf{9.31} & \textbf{13.14} & \textbf{35.80} & \textbf{6.31} & \textbf{9.16} & \textbf{13.06} & \textbf{35.87}  \\
\bottomrule
\end{tabular}
\label{tab:overall}
\end{table*}

In speaker domain adaptation, the speakers of the target domain may be a subset of the training set. Therefore, mapping all the target speech to source samples can be detrimental to adaptation. Following \cite{courty2017joint}, we adopt the Joint Distribution Optimal Transport (JDOT) framework with a joint cost between source and target data. As in \cite{lu2021unsupervised}, we use a soft weight with the \textit{Sigmoid} function and a bias $b$ to implement partial optimal transport. Additionally, we incorporate channel discrepancy into the cost function by adding the distance between source and target multi-scale convolutional features $h_{x_1}$ and $h_{x_2}$ in neural networks.

\begin{equation}
\begin{split}
C'(x_1, x_2) = \sigma( s * (C(y_{x_1}, y_{x_2}) & + \alpha_1 C(e_{x_1}, e_{x_2}) \\
& + \alpha_2 C(h_{x_1}, h_{x_2}) - b ))
\end{split}
\label{eq2}
\end{equation}
where $\sigma$ is the \textit{Sigmoid} function and $s$ is a scale factor. The multi-scale features are the output of the convolutional layers before the aggregation layers. 

Thus, the Joint Partial-OT function is reformulated as follows:

\begin{equation}\label{eq2}
\begin{split}
OT(u_1, u_2) & = \min_{(x_1, x_2) \in(u_1, u_2)} \sum {C'(x_1, x_2) \gamma(x_1, x_2)} 
\end{split}
\end{equation}

Finally, this joint partial-OT can be computed iteratively with the Sinkhorn algorithm \cite{cuturi2013sinkhorn}.

\subsection{Partial Regularized-OT Pseudo Label}

\begin{figure}[htb]
\begin{minipage}[b]{.45\linewidth}
  \centering
  \centerline{\includegraphics[width=\linewidth]{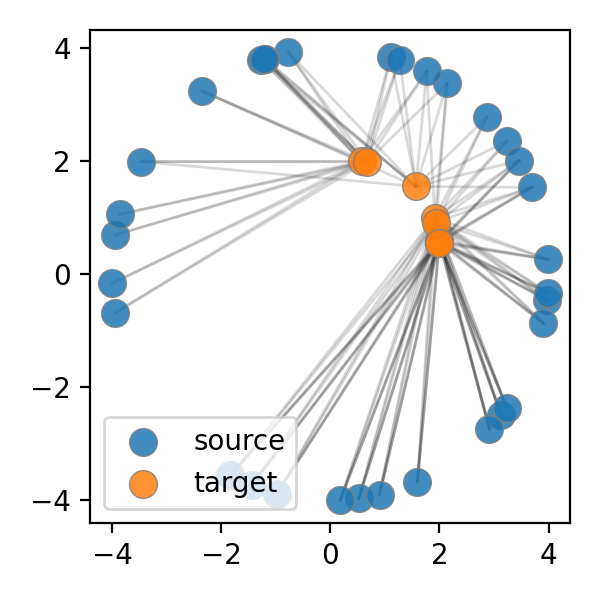}}
  \centerline{(a) Fully Unregularized OT}\medskip
\end{minipage}
\hfill
\begin{minipage}[b]{0.45\linewidth}
  \centering
  \centerline{\includegraphics[width=\linewidth]{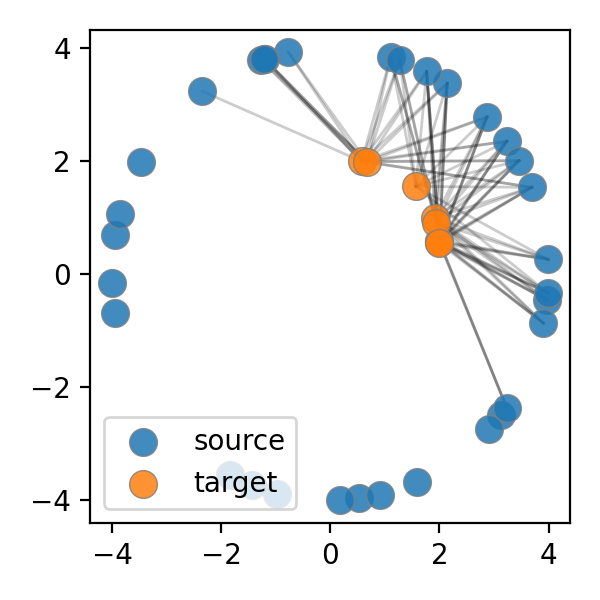}}
  \centerline{(b) Partial Regularized OT}\medskip
\end{minipage}
\caption{Unregularized / Regularized and Full / Parital Transport of Samples in Cosine Similarity Space.}
\label{fig:reg_ot}
\end{figure}

The optimal transport plan matrix can be naturally utilized for pseudo labeling. In speaker verification trained with margin-based cosine similarity loss \cite{deng2019arcface}, class prototypes ($p$) are explicitly embedded in the classifier. Computing the distance between target samples and class prototypes is straightforward, as this is the prediction of the source classifier. 

Here, we propose using the prediction (logits) as the cost function for OT-based pseudo-labeling. Additionally, considering the partial mapping between the target and single speech sample in the target domain typically contains only one speaker, the regularized OT is applied with an entropic regularization term, as in Figure~\ref{fig:reg_ot}. The transport plan can be calculated as:

\begin{equation}\label{eq2}
\begin{split}
\gamma'(x_2, p) &\triangleq \min_{\gamma \in(y_2)} \sum {c(y_2) \gamma'(x_2, p)} + \lambda \Omega(\gamma') \\
\Omega(\gamma') & = \sum \gamma'_{i,j} log(\gamma'_{i,j})
\end{split}
\end{equation}

Then, the indices $\hat{y_2}$ of the maximum values of the transport plan $\gamma'$ for the corresponding samples are selected as the pseudo labels. Given that the assignment of pseudo labels for direct classification can significantly impact model convergence, we opt for standard hard partial transport rather than assigning soft weights to samples. Specifically, we select samples with transport weights greater than the mean of the maximum weights within a batch.  To mitigate overfitting due to noisy labels, we optimize speaker classification using standard cross-entropy loss with a fixed temperature $\tau$, omitting the use of the original margin-based loss: 
\begin{equation}\label{eq2}
L_{pl}(x_2, \hat{y_2}) = \frac{1}{\sum{\delta_i}} \sum log \frac{\delta_i \cdot exp({cos(x_i, p_i) / \tau})}{\sum{exp({cos(x_i, p_j) / \tau})}}
\end{equation}

The weight $\delta_i$ is computed with the transport plan matrix $\gamma'$ as:
\begin{equation}
\delta_i = 
\begin{cases}
 1, &  \gamma'_i \geqslant \frac{1}{B}\sum\limits_{b} \max\limits_{r} \gamma'_{b,r} \\
 0, & \text{otherwise}
\end{cases}
\label{eq2}
\end{equation}
where the $B$ is the batch size, $b$ and $r$ is the number of row and column $\gamma'$.

\subsection{Unified Alignment with Pseudo Label}

In JPOT-PL, we integrate alignment and discriminative learning into a single framework, as illustrated in Figure \ref{fig:jpot}. The objective function can be expressed as:

\begin{equation}
L = L_{ce} + \eta L_{ot} + \beta L_{pl}
\label{eq2}
\end{equation}

\section{Experiement}
\label{sec:format}

\subsection{Settings}

\noindent \textbf{Dataset} VoxCeleb1 dataset \cite{nagrani2017voxceleb}, a large-scale English audio speaker dataset, is used. It comprises 1,211 speakers with 148,642 utterances for training and 40 speakers with 4,870 utterances for testing. We employ a radio corpus collection toolkit \cite{yang2024robust} to gather a Narrowband (3k Hz) radio corpus as the target domain data. Channel noise is inserted with energy levels of 0, 1, 2, and 5 volts for transmission channels.

\noindent \textbf{Model} The ECAPA-TDNN \cite{desplanques2020ecapa} is implemented within PyTorch. For model training, we use the same strategies in \cite{desplanques2020ecapa}. Input features are 80-dimensional Mel-frequency filter banks extracted from 2-second segments. Data augmentation is the same as in \textit{x-vectors} \cite{snyder2018x}. We adopt the AAM-Softmax loss \cite{deng2019arcface}  with a scale $s$ of 30 and a margin $m$ of 0.2. 

\noindent \textbf{Implement details} We select a subset of the radio VoxCeleb1 development set as target domain data, consisting of three subsets containing 64, 256, and 1211 (all) speakers, respectively. 

During testing, cosine similarity is computed on the Original Trials of the VoxCeleb1 test set, which contains 37,720 pairs. Model performance is evaluated using the Equal Error Rate (EER) averaged over 3 random seeds, with lower values indicating better performance.

\subsection{Results}

The overall results are presented in Table~\ref{tab:overall}, where we compare our method with other domain adaptation approaches in radio channel speaker verification scenarios. The EER in the source domain is the average of three adapting settings. The results show almost no performance degradation in the source domain. Except for DeepCORAL, other methods significantly reduce the EER in the target domains. The results demonstrate that our method (JPOT-PL) achieves the lowest EER in the target domain across various sizes of target adaptation sets, with over \textbf{10\%} (256 spk: 10.54\%$\rightarrow$9.31\%) reduction in some tests. As the adaptation data increases, the EER reduction becomes more pronounced, a trend not observed in other methods.

\begin{figure}[htb]
\begin{minipage}[b]{.45\linewidth}
  \centering
\centerline{\includegraphics[width=\linewidth]{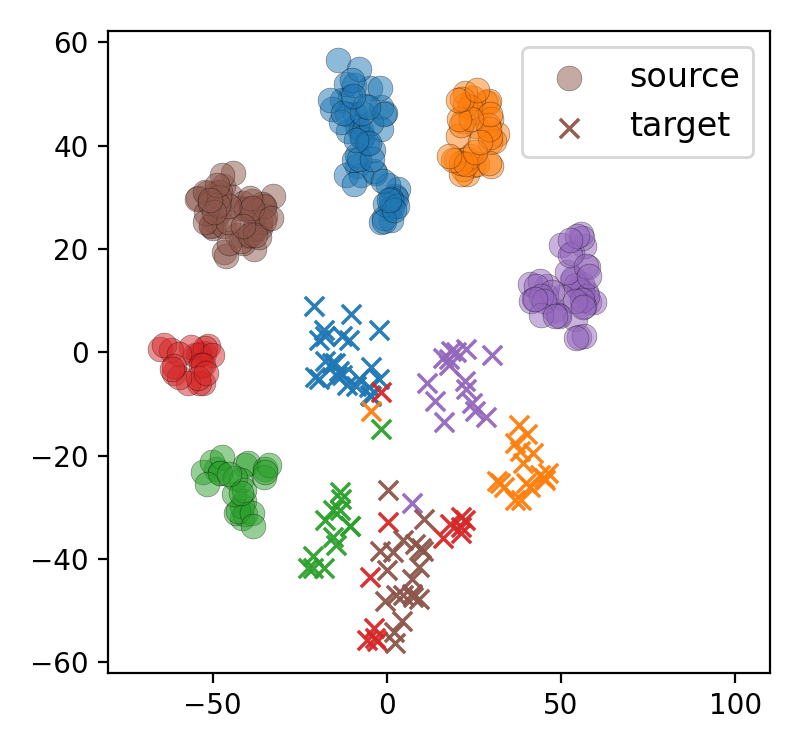}}

  \centerline{(a) No adaptation}\medskip
\end{minipage}
\hfill
\begin{minipage}[b]{0.45\linewidth}
  \centering
  \centerline{\includegraphics[width=\linewidth]{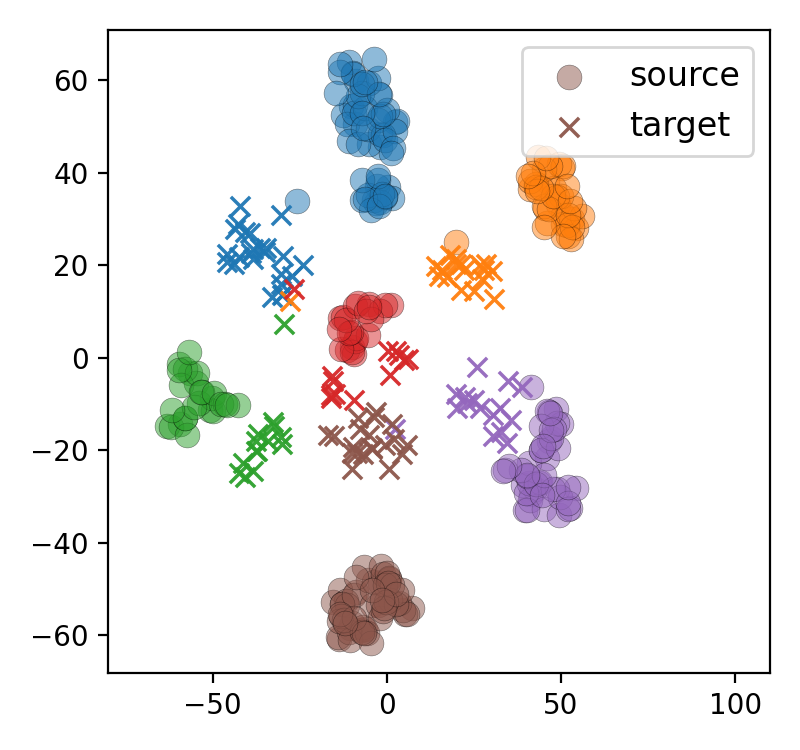}}
  \centerline{(c) DANN}\medskip
\end{minipage}
\begin{minipage}[b]{.45\linewidth}
  \centering
  \centerline{\includegraphics[width=\linewidth]{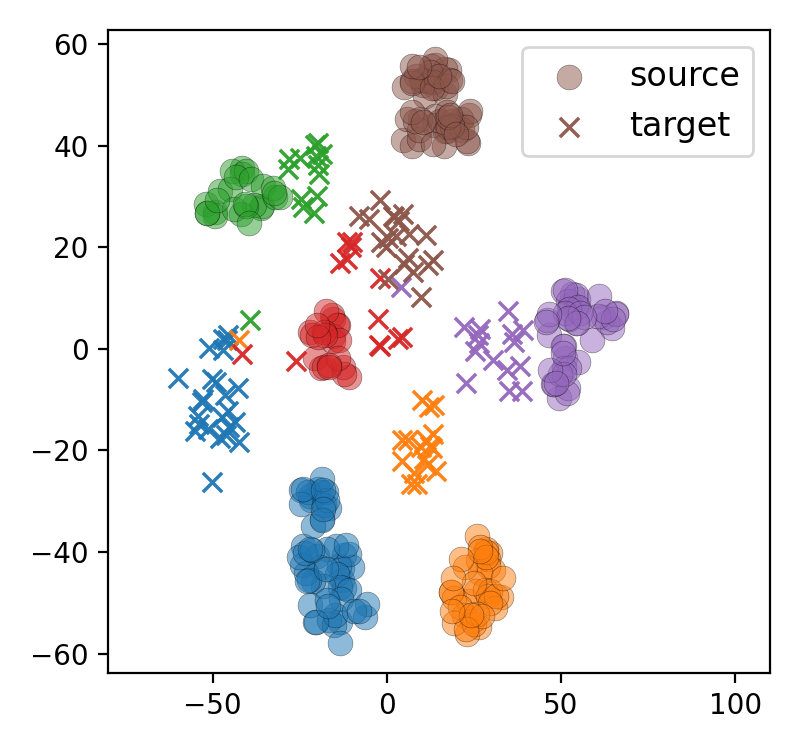}}
  \centerline{(d) DeepCORAL}\medskip
\end{minipage}
\hfill
\begin{minipage}[b]{0.45\linewidth}
  \centering
  \centerline{\includegraphics[width=\linewidth]{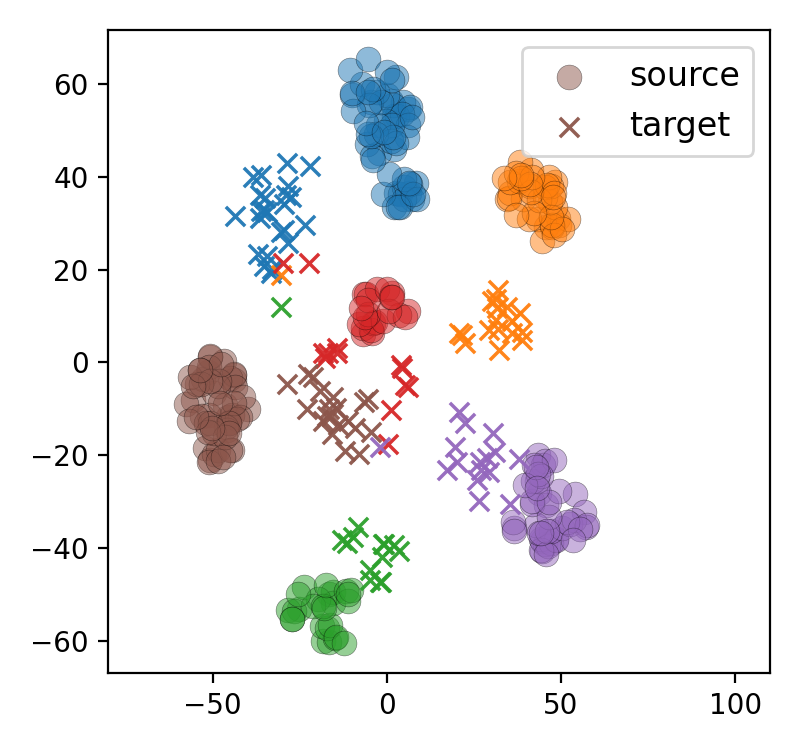}}
  \centerline{(e) MMD}\medskip
\end{minipage}
\begin{minipage}[b]{.45\linewidth}
  \centering
  \centerline{\includegraphics[width=\linewidth]{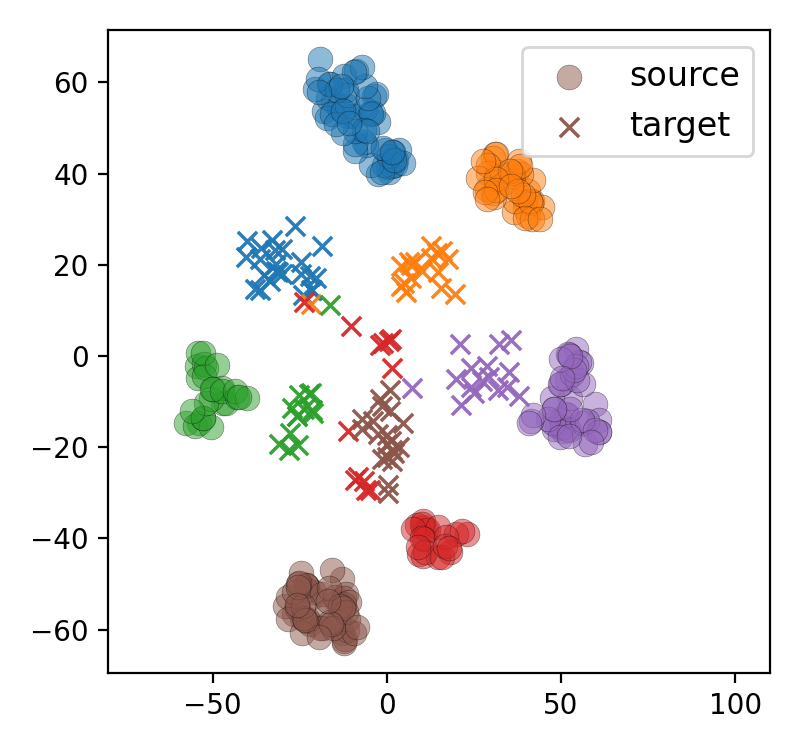}}
  \centerline{(f) DeepJDOT}\medskip
\end{minipage}
\hfill
\begin{minipage}[b]{0.45\linewidth}
  \centering
  \centerline{\includegraphics[width=\linewidth]{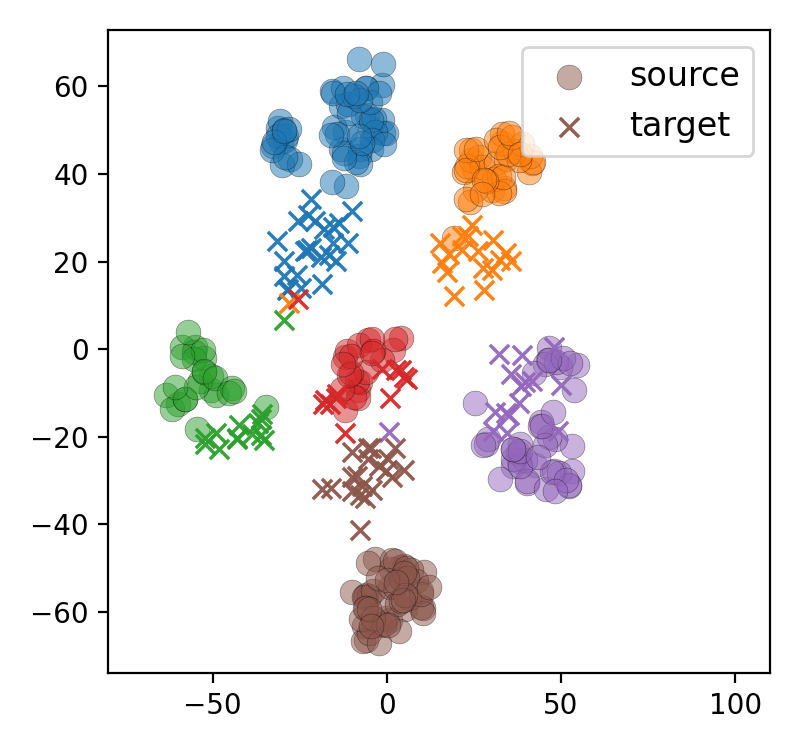}}
  \centerline{(g) JPOT-PL (Ours)}\medskip
\end{minipage}
\caption{T-SNE for speaker embeddings adaptation methods in VoxCeleb1 dev. 6 speakers are visualized with 6 colors.}
\label{fig:tsne}
\end{figure}

The t-SNE plots of speaker embeddings for DA methods are shown in Figure~\ref{fig:tsne}. In subfigure (a), without adaptation, the embeddings for the target domain are distorted and mixed with source domain ones. All DA methods bring the target samples closer to the corresponding clean source samples. However, our method achieves superior adaptation results with enhanced alignment, especially for the speaker embeddings in orange, where other methods fall short.

\subsection{Ablation Studies}

\subsubsection{Alignment and Pseudo Label}

The results of ablation studies for the two components of JPOT-PL are presented in Table~\ref{tab:hyper}. Here, $\eta$ represents the importance of JPOT alignment loss, and $\beta$ denotes the weight of pseudo-label classification. 

\begin{table}[h]
\centering
\caption{Performance ( EER($\%$) ) of ablation studies for two components in JPOT-PL. [0,1,2,5]: noise volts in channels.}
\begin{tabular}{llllllllll}
\toprule
\multicolumn{1}{c|}{\multirow{2}{*}{\textbf{Components}}} & \multicolumn{1}{c}{\multirow{2}{*}{\textbf{Value}}}  & \multicolumn{4}{|c|}{\textbf{256 Spks}}       \\
\multicolumn{1}{c|}{}                               &         & \multicolumn{1}{|c}{0} & \multicolumn{1}{c}{1} & \multicolumn{1}{c}{2}     & \multicolumn{1}{c|}{5}  \\
\midrule
-                                     &    -   & 8.15                  & 12.60                 & 16.68  & 39.05                            \\
\midrule
+JPOT                                    &  $\eta$=1, $\beta$=0       &  7.10 & 10.39 & 14.26 & 37.61                            \\
+PROT-PL                                    &    $\eta$=0, $\beta$=0.1   &  6.58 & 9.51 & 13.25 & \textbf{35.83}                           \\
\midrule
JPOT-PL                                    & $\eta$=1, $\beta$=0.1    & \textbf{6.48} & \textbf{9.41} & \textbf{13.22} & 35.86 \\
\bottomrule
\end{tabular}
\label{tab:hyper}
\end{table}

\begin{table}[h]
\centering
\caption{Top-K Classification Accuracy (\%) with two Pseudo Lable metrics. \textbf{$\gamma'$-PROT} chooses high confidence (50\%) samples. \textbf{bold}: the best accuracy, \underline{underline}: the second best accuracy.}
\begin{tabular}{lcrrrr}

\toprule
\multicolumn{1}{c|}{\multirow{2}{*}{\textbf{Metric}}} & \multirow{2}{*}{\textbf{Epoch}} & \multicolumn{2}{|c|}{\textbf{Source}} & \multicolumn{2}{|c|}{\textbf{Target}}       \\
\multicolumn{1}{c|}{}                         &         & \multicolumn{1}{|c}{top-1}  & \multicolumn{1}{c|}{top-5} & \multicolumn{1}{|c}{top-1}  & \multicolumn{1}{c|}{top-5}  \\
\midrule
logits   & \multirow{4}{*}{0} & 99.54 & 99.83 & 37.10 & 58.55   \\
$\gamma'$-OT &   &  46.50 & 74.87  & 27.06 & 48.15 \\
$\gamma'$-ROT &   &  99.24 & 99.74  & \underline{49.03} & \underline{68.76} \\
$\gamma'$-PROT &   &  100.00 & 100.00  & \textbf{77.30} & \textbf{90.54} \\
\midrule
logits  & \multirow{4}{*}{1}   &  99.57 & 99.85 & \underline{63.69} & \underline{86.97}  \\
$\gamma'$-OT &   &  47.15 & 75.06  & 28.32 & 49.42 \\
$\gamma'$-ROT &    &  99.21 & 99.76  & 57.88 & 77.01 \\
$\gamma'$-PROT &    & 100.00  & 100.00  & \textbf{85.02} & \textbf{93.93} \\
\midrule
logits  & \multirow{4}{*}{2}  &  99.56 & 99.84 & \underline{63.43} & \underline{82.23}   \\
$\gamma'$-OT &    &  46.80 & 74.90  & 28.58 & 49.56 \\
$\gamma'$-ROT &    &  99.25 & 99.76  & 58.19 & 77.36 \\
$\gamma'$-PROT &   &  100.00 & 100.00  & \textbf{84.27} & \textbf{93.68} \\
\bottomrule
\end{tabular}
\label{tab:metric}
\end{table}

As shown in Table \ref{tab:hyper}, both Joint Partial OT alignment and Partial Regularized-OT Pseudo Labeling contribute to improved domain adaptation results. The combination of these two components leads to further performance improvement.

\subsubsection{Pseudo Metric} 
We perform pseudo-label classification during domain adaptation. In addition to using $\gamma'$-PROT in JPOT-PL, we also examine the logits, $\gamma'$-OT from standard OT, and $\gamma'$-ROT from regularized OT. The top-K classification accuracy on the target domain data is shown in Table~\ref{tab:metric}.

We calculated the accuracy of the JPOT-PL models during adaptation. Pseudo labeling with $\gamma'$-PROT achieves significantly higher accuracy at the beginning of the adaptation process (Epoch 0). After alignment, the accuracy of logit labels and $\gamma'$-ROT labels becomes similar on the target domain. 

\section{Conclusion}
\label{sec:pagestyle}

In this paper, we propose a domain adaptation method for speaker verification in degraded channels. Building on Optimal Transport, we introduce Joint Partial-OT for domain alignment and Partial Regularized-OT for pseudo labeling. To address partial alignment and channel variabilities, we combine channel-related multi-scale features with speaker embeddings for joint partial-OT domain distance computation. Furthermore, by incorporating pseudo labels with Partial Regularized-OT for classification, we enhance the discriminative ability in the target domain after adaptation. Our experiments on radio channel conditions were conducted using the VoxCeleb dataset. Compared with several state-of-the-art domain adaptation methods, the results demonstrate the effectiveness of our approach. Some methods fail in this scenario due to misalignment or a lack of a discriminative learning process. However, severe channel degradation still poses challenges in speaker verification, warranting further research.

\bibliographystyle{IEEEtran}
\bibliography{refs}

\vspace{12pt}
\color{red}

\end{document}